\documentclass[aip,jcp,amsmath,amssymb,reprint,numeric]{revtex4-1}
\usepackage{graphicx}
\usepackage{subfigure}
\usepackage{color}
\usepackage{amsmath}
\usepackage{bm}

\begin{document}
\title {Comparitive study of anomalous size dependence of charged and neutral solute diffusion in water}
\author{Sayantan Acharya}
\author{Sarika Maitra Bhattacharyya}
\email{mb.sarika@ncl.res.in}
\affiliation{\textit{Polymer Science and Engineering Division, CSIR-National Chemical Laboratory, Pune-411008, India}}

\date{\today}


\begin{abstract}
In this work, we perform a comparative study of the size dependence of diffusion of charged and neutral solutes in water. 
The neutral solute in water shows a nonmonotonicity in the size dependence of diffusion. This is usually connected to the well known Levitation effect where it is found that when solute diffuses through the transient solvent cages 
then for attractive solute-solvent interaction and for a particular size of the solute there is a force balance which 
leads to the maximum in diffusion. 
Similar maximum in diffusion of charged solutes has also been observed and connected to Levitation effect. However, earlier studies of ionic diffusion 
connects this nonmonotonicity to the interplay between hard sphere repulsion and 
Coulombic attraction. In this work, we show that although the size dependence of both charged and neutral solutes have 
a nonmonotonicity, there is a stark difference in their behaviour. For charged solute with increase in attraction the 
maximum shifts to higher solute sizes and has a lower value whereas for neutral solute it remains at the same place and 
has a higher value. 
We show by studying the ionic and non-ionic part of the potential that for larger solutes it is the nonionic part which dominates and for smaller solutes the ionic part and the is a transition between them. As the charge on the solute increases, this transition takes place at larger solute sizes which leads to the shift in the diffusivity maxima and reduction of the peak value. 
We show that although the charged solutes also explore the solvent cage even before we reach the size which Levitates due to Coulombic 
attraction the diffusion value drops. Thus the origin of diffusivity maxima in charged and neutral solute diffusion is different. 
\end{abstract}
\maketitle

\clearpage

\section{Introduction}

Anomalous diffusion of ions in water or in any other polar solvent  is a long-standing problem \cite{Born1920,size,strc1,strc2}.
According to Walden’s rule, ionic conductivity, i.e., diffusion of ions should be inversely proportional to the ion radius \cite{walden,waldenrule}.  
However, in polar solvents, this relation is not followed. There is a breakdown in the linear behaviour and a size dependent peak in the conductivity is observed.  
Over the years, to explain this anomaly several theories have been put forward by different groups \cite{Born1920,chen,boyd,onsager,zwanzig1,zwanzig2,wolynes,rasaiah,biswas,ranjit,bagchi,evidence,size,strc2,padma}. 

In the continuum picture\cite{Born1920,chen,boyd,onsager,zwanzig1,zwanzig2,ranjit,bagchi,biswas,rasaiah,wolynes}, the friction on the ion has a viscous and a dielectric part.
The dielectric friction is higher for smaller sizes whereas the viscous friction for larger sizes. An interplay between the two terms give rise 
to a diffusivity maximum. However, the dielectric friction part is symmetrical w.r.t the sign of the charge. Thus these theories could not explain the difference
between the positive and the negative ions. The difference between these ions is believed to arise due to the difference in the solvent structure around them.
Chen and Adelman\cite{chen} in their extension of the continuum model included the effect of the local structure of the solvent. However, instead of dielectric
friction, they considered only the viscous mode with an effective radius which depends on the size of the bare ion, the solvated ion and also on the 
degree of solvation. They could show that for small sizes, the degree of solvation is high, which gave rise to a large effective radius. However, for intermediate
sizes the degree of solvation is less, which gave rise to a small effective radius and thus high diffusion value. But, in 
this theory, the separation between dielectric and hydrodynamic friction is not clearly defined. 

There have also been molecular theories like that of
Wolynes\cite{wolynes} where the friction on an ion was separated into that arising from the soft and hard part of the potential. The theory in certain 
limits reproduced the continuum picture.   

Later, Bagchi and co-workers\cite{bagchi,biswas} have extended this approach by including the intermolecular
orientational correlations of the solvent as well as the self-motion of the ion. The results are in excellent agreement with the experimental studies. According to their study, the diffusivity maxima for certain solute sizes arises due to cancellation between hard-sphere and electrostatic part of the interaction.

In a computer simulation (MD) study of alkali ions and halide ions in water\cite{rasaiah,lyndenbell} all the different theories and their approximations were tested. Chandra and coworkers, through a simulation study, tried to connect this size dependency of diffusivity maxima with hydrogen bonding \cite{chandra}. 

Ghorai and Yashonath\cite{evidence} have studied the size dependence of the diffusivity of charged solutes in water where they have systematically varied the size and also performed the study for different values of the charge both for positive and negative ions. They have shown that above certain value of the charge, the system shows a diffusivity maximum which they connected to the Levitation Effect (LE) obtained in their earlier studies on neutral systems \cite{padma,kumar,strc2,size,strc1}.
In those studies on neutral systems they have shown that small solutes move through the transient cages formed by the solvent. For attractive solute-solvent interaction the solute while passing through the neck of the cage feels an attraction towards one side of the neck and thus gets stuck which gives rise to reduced diffusion. When the size of the solute is about 80$\%$ of the size of the neck, the attractive forces from all directions are equal and in opposite direction causing a force balance. This allows the solute to freely pass through the neck without getting attached to the wall. The authors have claimed that this force balance is universal and is also observed in ionic systems \cite{strc1,strc2,size}. Note that, except for the study of Ghorai and Yashonath\cite{evidence} all the other studies although different in some aspect \cite{wolynes,rasaiah,ranjit,biswas}, explained the diffusivity maxima in ionic diffusion in terms of interplay between short range hard sphere and long range ionic interactions.

 In order to understand the connection between LE and diffusivity maxima in ionic systems, we present a study of diffusion of both charged and neutral solutes in water. Similar to Ghorai and Yashonath\cite{evidence}, we systematically vary the size and also perform the study for multiple values of the charge. We find that, unlike for non-ionic solutes, where the diffusivity maxima remains at the same position as the interaction between the solute and solvent is increased, for an ionic solute it shifts to larger sizes as the ionic charge and thus the interaction increases. We also find that the peak height of the diffusivity maxima reduces with increase in the interaction which is just the opposite to that obtained for non-ionic solutes. Here, we explain the origin of these differences. Our observations are quite similar to that reported by Ghorai and Yashonath \cite{evidence} but our analysis and interpretations are quite different.
 We find that
 although the solute particle diffuses through the solvent cage the diffusivity maxima does not arise due to force balance for a particular size of the solute but it arises due to the interplay between the Lenard Jones and ionic interactions.

The next section contains Computational details. Section 3 includes the results and 
discussion followed by the conclusion in section 4.

\section{Computational Details}
\subsection{Intermolecular Potential Functions}

\textit{Water-Water.} We consider SPC/E water model \cite{spce,spce2} in our simulations. This is a three point water structure. Three sites representing one Oxygen (O) and two Hydrogen (H) atoms. 
The O-H bond length is 1 \AA{}. The HOH angle is 109.47\textordmasculine. Charge of individual O atom is given as -0.8476e and H atom is given as +0.4238e. 
A short range Lennard-Jones (LJ) potential along with a long range Coulomb potential makes the whole equation look like,

\begin{equation}
\Phi_{ww}=4\epsilon_{OO}[(\frac{\sigma_{OO}}{r_{OO}})^{12}-(\frac{\sigma_{OO}}{r_{OO}})^{6}]+\Sigma_{i\neq j}\frac{q_{i}q_{j}}{r_{ij}},
\end{equation}  
here, $\epsilon_{OO}$ and $\sigma_{OO}$ are LJ parameters between Oxygens of two water molecules and are defined in this model as 0.650 kJ/mol and 3.166 \AA{} respectively. $r_{OO}$ is the distance between them. The charge at site i is $q_{i}$.

\textit{Solute-Solute.} The interaction between two solute particles is considered as a sum of short range LJ and long range Coulomb potential. This is expressed as, 

\begin{equation}
	\Phi_{ss}=4\epsilon_{ss}[(\frac{\sigma_{ss}}{r_{ss}})^{12}-(\frac{\sigma_{ss}}{r_{ss}})^{6}]+\frac{q_{s}q_{s}}{r_{ss}},
\end{equation}

Here, the $\sigma_{ss}$=1.5 \AA{} and $\epsilon_{ss}$=0.2608 kJ/mol are fixed. We consider the charge on the solute as, $q_{s}=0, \pm0.001e, \pm0.01e, \pm0.05e$ and $\pm0.3e$. 

\textit{Solute-Water.} We consider the solutes to be charged spheres. It has short range LJ interaction with water Oxygen and long range Coulombic interaction with both Oxygen and Hydrogen atoms of a water molecule. The form of the potential looks like,

\begin{equation}
\Phi_{sw}=4\epsilon_{sO}[(\frac{\sigma_{sO}}{r_{sO}})^{12}-(\frac{\sigma_{sO}}{r_{sO}})^{6}]+\Sigma_{j\ne s}\frac{q_{s}q_{j}}{r_{sj}},
\end{equation} 
Where, the interaction strength is $\epsilon_{sO}$=1.5846 kJ/mol. For neutral solutes, we increase this strength by 10 and 15 times for two different set of systems. 
 We allow interpenetration between solute and water, so the $\sigma_{sO}$ does not obey the Lorentz-Berthelot combination rule. We keep the $\sigma_{ss}$ fixed but vary $\sigma_{sO}$ by varying the interpenetration.
  We take a range of  $\sigma_{sO}$ values as, from 0.9 \AA{} to 1.3 \AA{} with a gap of 0.1 \AA{}, from 1.3 \AA{} to 2.5 \AA{} with a gap of 0.2 \AA{} and 3.0 \AA{}.
While choosing the range of radius, we make sure that at small distance, the repulsive part of the LJ interaction between the solute (SOL) and Oxygen atom of water dominates 
over the Coulombic interaction. This is specially important in case of -ve charges as for small sizes, the Coulombic interaction between the -ve charge 
and the Hydrogen atom of water can be very strong and this will result in the -ve charge sitting on the H atom. So, for q=-0.01e and -0.05e, the minimum $\sigma_{sO}$ value is 1.3 \AA{}, while for q=-0.3e the minimum $\sigma_{sO}$ value is 1.7 \AA{}.

However, note that, for +ve charges this issue doesn't arise as at small distance, it is always the repulsive part of the LJ potential which dominates. 
But since, the simulations are done with equal number of +ve and -ve charges, we keep the range similar for both the charges.

\subsection{Simulations Details}
\label{sim}

  We perform Molecular Dynamics (MD) simulations using GROMACS  package \cite{gromacs1,gromacs2}.
 We take a system with 2597 solutes where 44 of them are ions surrounded by 851 water molecules. Among the 44 ions, half of them are positive ions and rest are
 negative ions. 

We use the isothermal-isobaric ensemble (NPT) simulation for equilibration run at T=300 K and at reduced pressure of 0.7 Bar. The production run is done in microcanonical (NVE) ensemble. 
 
 The MD simulations are performed in a cubic box  using Nos\'{e}-Hoover thermostat\cite{nose} and Berendsen barostat. The integration 
step is varied for different charges depending on the strength of the ion-water interaction. We have a range of integration steps from 0.6 fs to 0.0002 fs. For smaller sizes and higher charges, integration steps are smaller. 
 In this study, length and temperature 
are given in real units.  
 All the above mentioned systems are equilibrated for 150-300 ps followed by a production 
run of 800 ps. Systems with higher charges are equilibrated over longer times.

\subsection{Methodology}

In this work, we calculate the diffusion coefficient, D from both mean-square displacement (MSD)  and and velocity autocorrelation function (vacf). We calculate the MSD as,

\begin{equation} 
\langle \Delta r^{2}(t)\rangle=\frac{1}{N}\sum_{i}\langle(r_{i}(t)-r_{i}(0))^{2}\rangle ,
\end{equation}
\noindent
where, $r_{i}(t)$ is the position at time t and N is the number of particles. From the long time behaviour of MSD, the diffusion coefficient D can be written as,

\begin{equation}
D= \lim_{t\rightarrow \infty} \frac{\langle \Delta r^{2}(t)\rangle}{6t}
\end{equation}
\noindent
At longer time by fitting the MSD with time, we obtain D from the slope of the fitted plot. 
 
The diffusion value can also be obtained from vacf as, 

\begin{equation}
D=\frac{1}{3}\int_{0}^{\infty}dt\langle v_{i}(t)\cdot v_{i}(0)\rangle ,
\end{equation} 
\noindent
where, $v_{i}(t)$ is the centre-of-mass velocity of a single molecule at time t.  

Self-diffusion is often described in terms of the Stokes-Einstein (SE) relation\cite{brown1,brown2}. The equation predicts an inverse dependence of the solute diffusion, D, on the solvent viscosity, $\eta$, and solute radius $\sigma_{ss}$. The expression is written as,

\begin{equation}
D=\frac{k_{B}T}{s\pi\eta \sigma_{ss}},
\end{equation}
\noindent
where, s = 4 for slip boundary condition and  6 for stick boundary condition. However, this relation is found to be violated for small solutes\cite{smb,kadar,sayantan} and also in supercooled liquids\cite{supercool,marcus,sill}.


\section{Results and Discussion}

In this work, we study the nonmonotonicity of diffusion as a function of solute size. Although the primary focus of this work 
is to study the diffusion of charged solutes (ions) in water, for the sake of comparison, we first present a study of the diffusion of neutral solutes in water. 

\subsection{Neutral Solutes}
Yashonath and coworkers in their study of diffusion of neutral solutes in water, interacting only via LJ interaction, at T=180K,
have shown that for solute-solvent interaction $\epsilon_{sO}=1.5846$ kJ/mol there is a nonmonotonicity in the size dependence of diffusion \cite{ghorai}. However, in that case, due to low temperature, the water dynamics was almost frozen.  
In this present study, we focus on room temperature water dynamics, i.e, T=300K. At room temperature, with solute-solvent interaction 1.5 kJ/mol the 
nonmonotonicity in size dependence of diffusion disappears \cite{temp,strc2}. This is because the kinetic energy of the solute at room temperature is large enough to overcome the attraction. Thus, to obtain a nonmonotonicity in solute diffusion, we  arbitrarily increase the $\epsilon_{sO}$  value to 15 and 22.5 kJ/mol, which is about 10 and 15 times the value used in the earlier study \cite{ghorai}.

 As shown in Fig.1a, the diffusion as a function of 1/$\sigma_{sO}$ does show a nonmonotonic behaviour with peak at $\sigma_{sO}\backsimeq$1. Note that, the position of the peak is similar to that obtained earlier\cite{ghorai2,manoj,sayantan}. The presence of such nonmonotonicity can be attributed to LE, i.e, for solute-solvent size $\sigma_{sO}\simeq$1 the solute, while passing through the neck of the transient solvent cage does not feel strong attraction due to force balance and this leads to the increase in the diffusion value. As discussed earlier \cite{size,ghorai2,ghorai,Bhatia,manoj,santikary} this force balance is specific to the size of the solute as compared to the size of the neck of the cage. 
  The observation made from the present study is that at the point of levitation, where there is expected to be a force balance, the diffusion value is almost independent of interaction. We also find that, the degree of levitation, i.e., the comparative height of the maximum in self-diffusivity increases with increase in solute-solvent interaction, which has been reported earlier in other systems \cite{Bhatia,size,manoj,sayantan}. 

Note that, compared to that of a standard LJ system the position of the peak appears at a smaller $\sigma_{sO}$ value. This can be related to the water structure and thus the size distribution of the transient cages and their necks. 

\subsection{Charged Solutes}
Next we study the diffusion of charged solute particles in water. To understand the effect of the charge on solute diffusion we first study the uncharged solute and then slowly increase the charge on the solute and study its diffusion. The range of the solute sizes studied varies with the charge to avoid some unphysical systems. The details of the chosen range are given in Sec II. 
 
 \begin{figure}[ht]
 	\centering
 	\subfigure{
 		\includegraphics[width=0.36\textwidth]{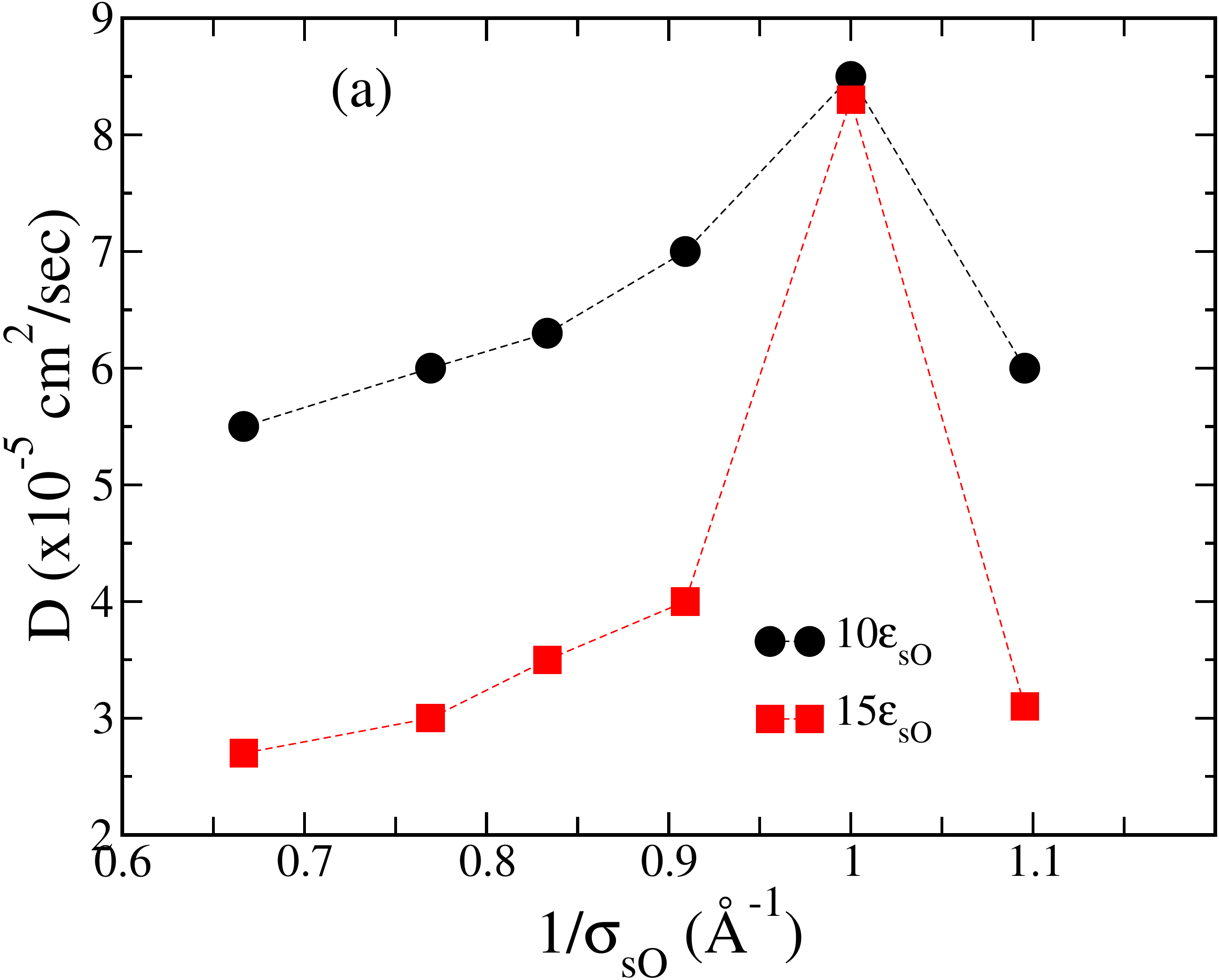}}
 	\subfigure{
 		\includegraphics[width=0.36\textwidth]{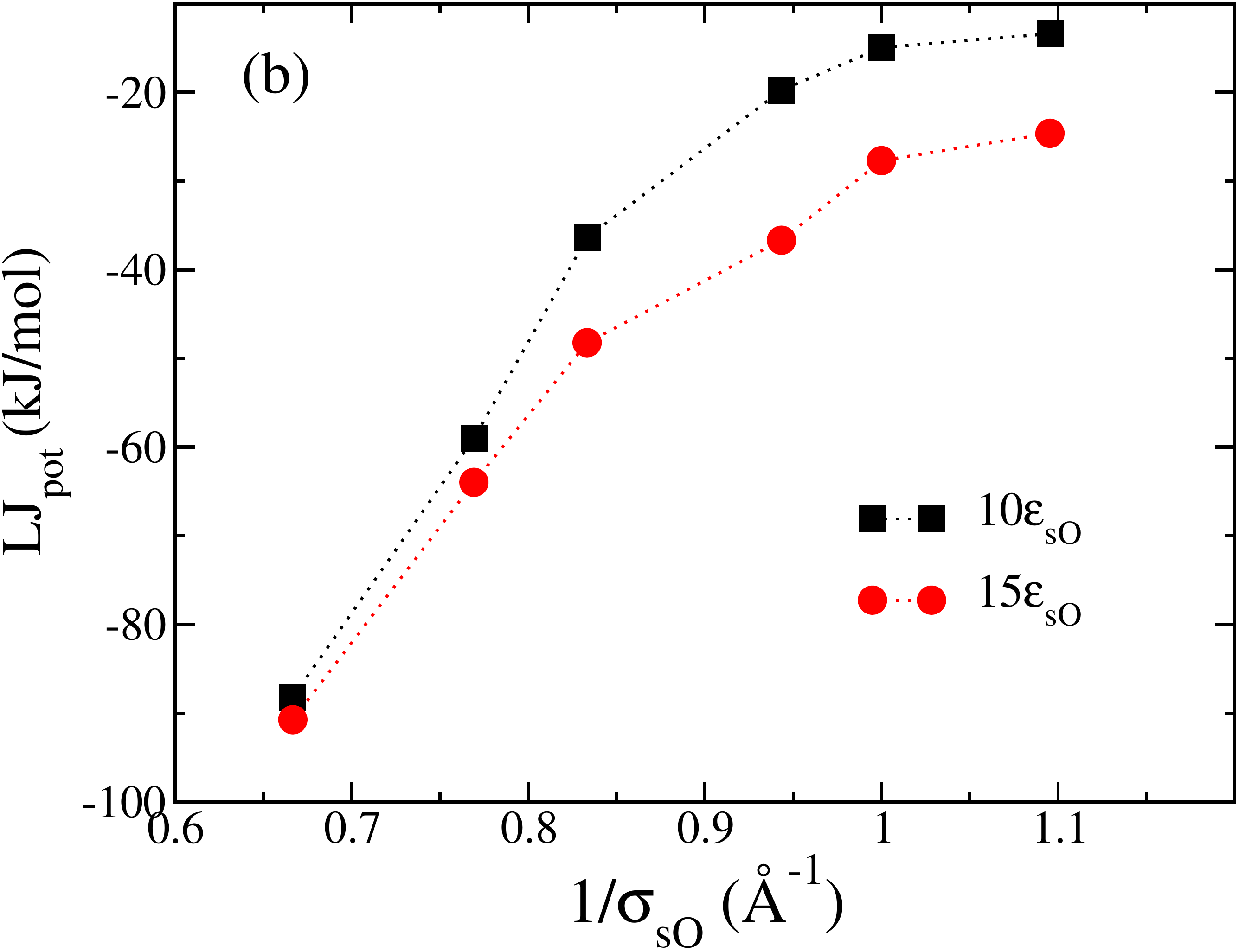}}
 	
 	\caption{\it{(a) The diffusion coefficient D as a function of 1/$\sigma_{sO}$ for neutral solutes in water. The solute-solvent interaction is kept very high (10 and 15 times $\epsilon_{sO}$ for black square and red circle respectively). (b) LJ potential of these two systems as a function of 1/$\sigma_{sO}$. The dotted lines are guides to the eye.}}
 	\label{fig1}
 \end{figure}
 \noindent

\begin{figure}[ht]
	\centering
	\subfigure{
		\includegraphics[width=0.36\textwidth]{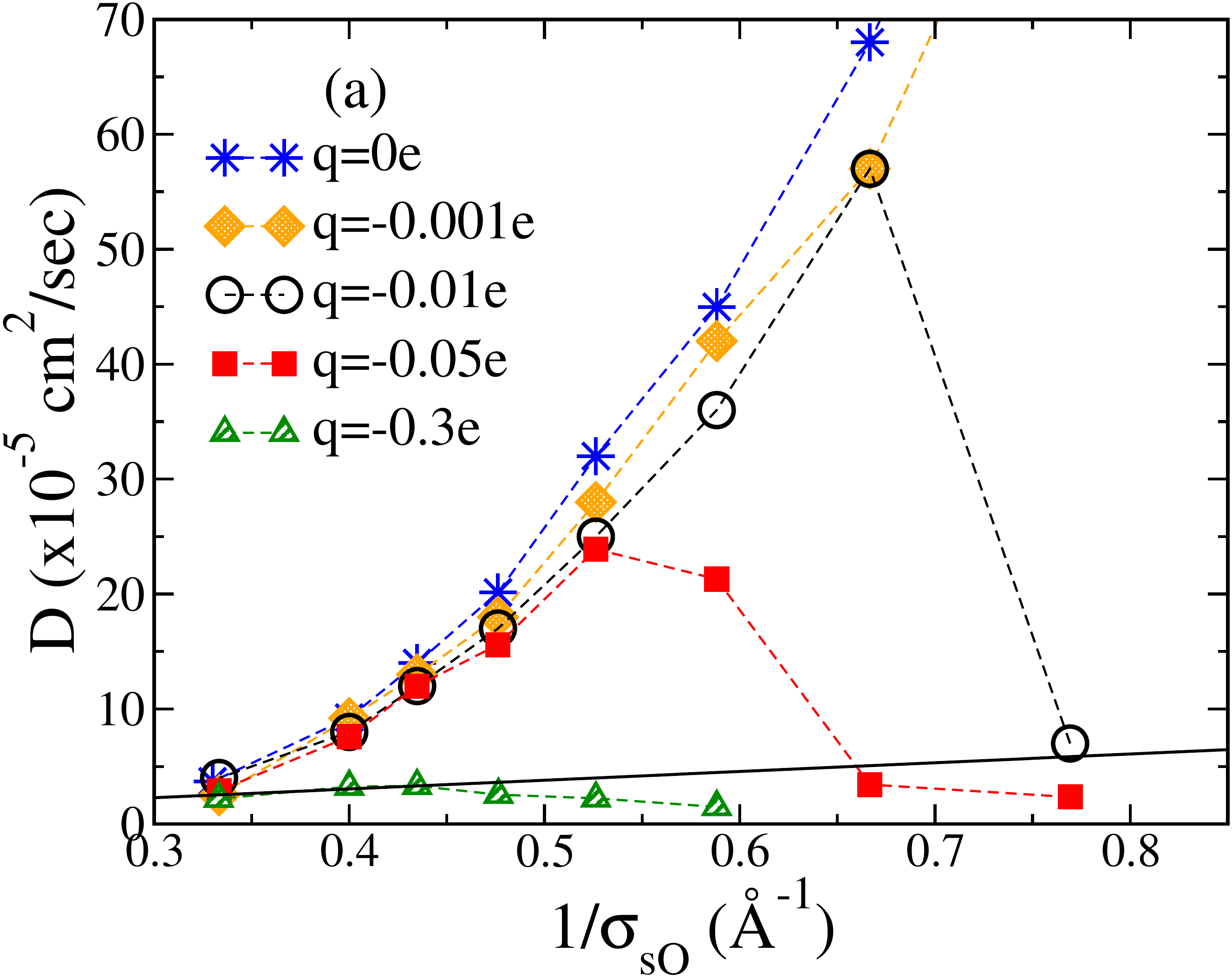}}
	\subfigure{
		\includegraphics[width=0.36\textwidth]{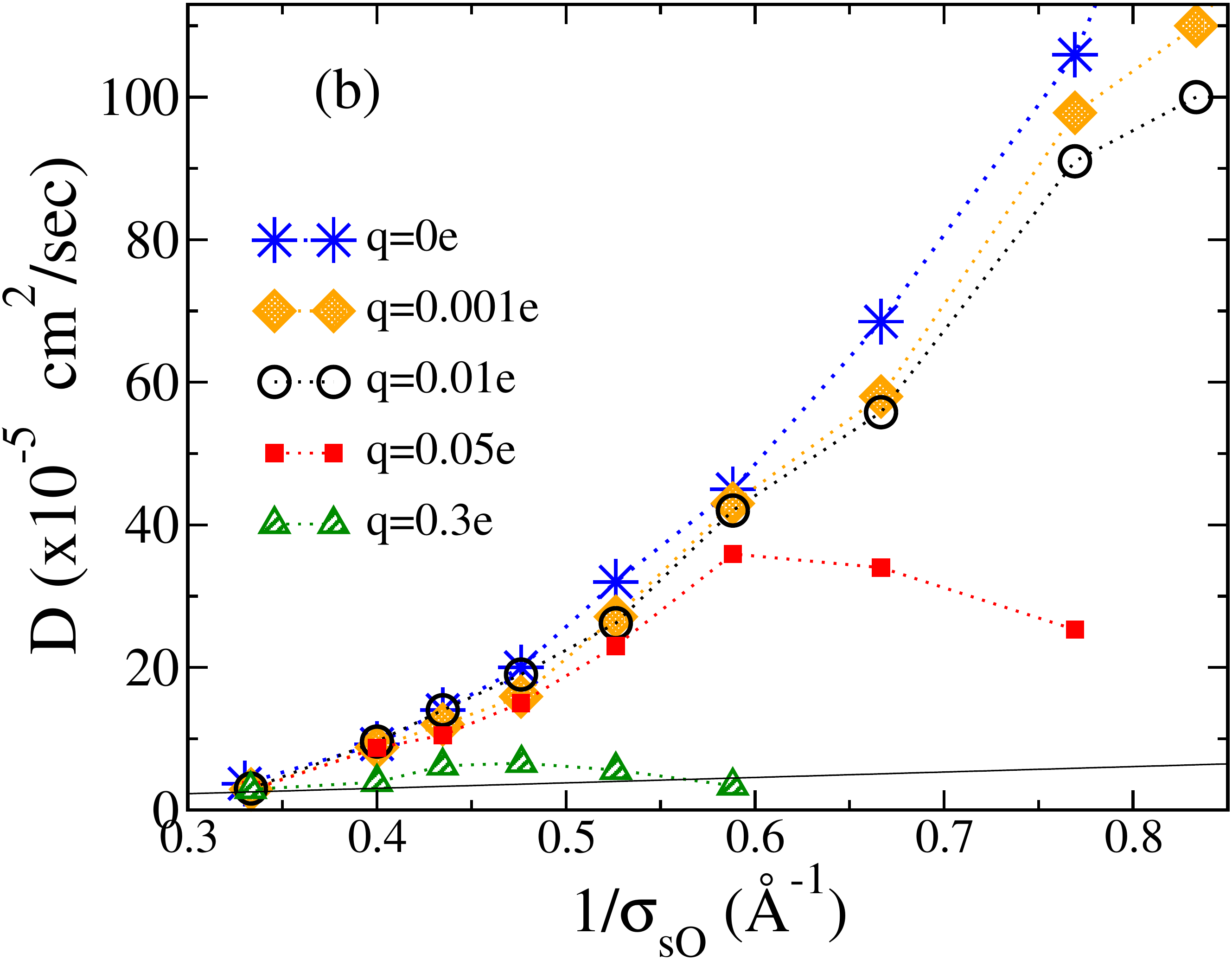}}
	\caption{ \it{ The diffusion coefficient D as a function of 1/$\sigma_{sO}$ for neutral and charged solutes in water. (a) For negatively charged solutes the diffusivity maxima is observed for $q_{s}$=-0.01e, -0.05e and -0.3e. (b) For positive ions the diffusivity maxima is observed for $q_{s}$=-0.05e and -0.3e. Rest of the charges show almost linear behaviour like the neutral solute. Solid line is the SE prediction. The dotted lines are guides to the eye. }}
	\label{fig2}
\end{figure}
\noindent

Here we choose $\epsilon_{sO}=1.5$ kJ/mol. When the solute is neutral this system does not show any nonmonotonicity at T=300K. Thus, we know that the LJ interaction alone is not strong enough to produce a diffusivity maximum.
In Fig.2 we plot the diffusion values as a function of inverse $\sigma_{sO}$ for different charges. Our observation is similar to that reported by Ghorai and Yashonath \cite{evidence}. However, the range of charges studied here is different. For smaller charges both for the negatively and positively charged solutes we do not observe any nonmonotonicity. For +ve charges the nonmonotonicity appears for $q_{s}\geq$+0.05e and for -ve charges it appears for $q_{s}\geq$-0.01e. We find that as we increase the charge on the solute, there 
 is a shift of the diffusivity maxima towards bigger sized solutes.
For example, we see that the maxima is near $\sigma_{sO}$=1.5 \AA{} for $q_{s}$=-0.01e, whereas, for $q_{s}$=-0.3e the maxima is near $\sigma_{sO}$=2.3 \AA{}. 
An  increase in the magnitude of '$q_{s}$' reduces the height of the maxima in the diffusion plot. These findings are similar to that obtained by Ghorai and Yashonath \cite{evidence}. The only difference is that, they claimed that  the nonmonotonicity is maximum for $q_{s}$=0.1e, whereas we find that it exists even at lower values of '$q_{s}$'. 
For positive charge at $q_{s}$=0.01e the diffusivity maximum disappears. However, for this system the diffusivity maxima appears at lower temperatures (not shown here). Thus, we find that the appearance of the diffusivity maxima is dependent on multiple factors. It not only depends on the magnitude of the charge but also on the sign of the charge and the temperature of the system. 
The dependence on the sign tells us that the water structure near a +ve and -ve charge is different, which has been reported earlier \cite{barbara,marcus}. 

If we now compare the nonmonotonicity as observed for charged (Fig.2a) and neutral (Fig.1a) solutes, we find that, (i) for charged solute with increase in attraction the peak of the diffusivity maxima shifts to larger sizes which in case of neutral solute remains fixed, (ii) for charged solutes the diffusivity maxima decreases with increase in interaction whereas the opposite trend is observed for neutral solutes. 
Thus although, both charged and uncharged solutes show a nonmonotonicity in size dependence of diffusion, there exist certain stark differences in the nature of the nonmonotonicity as discussed in earlier studies \cite{evidence} and also presented here. The rest of the article is devoted to understand the origin of this difference.

\subsubsection{ Shift of the position of diffusivity maxima }

First, we address the shift of the diffusivity maxima to larger solute-solvent radius with the increase in solute charge. Note that in our study the size dependence is incorporated by changing $\sigma_{sO}$. The radius of the solute remains unchanged.

As discussed earlier in LE, the presence of the diffusivity maxima is intimately connected to the solvent transient cage formation and the solute exploring the cage diffusion \cite{strc2,sayantan}. Thus the position of the diffusivity maxima in LE provides us the information of the neck size of the transient cage.

\begin{figure}[ht]
	\centering
	\subfigure{
		\includegraphics[width=0.36\textwidth]{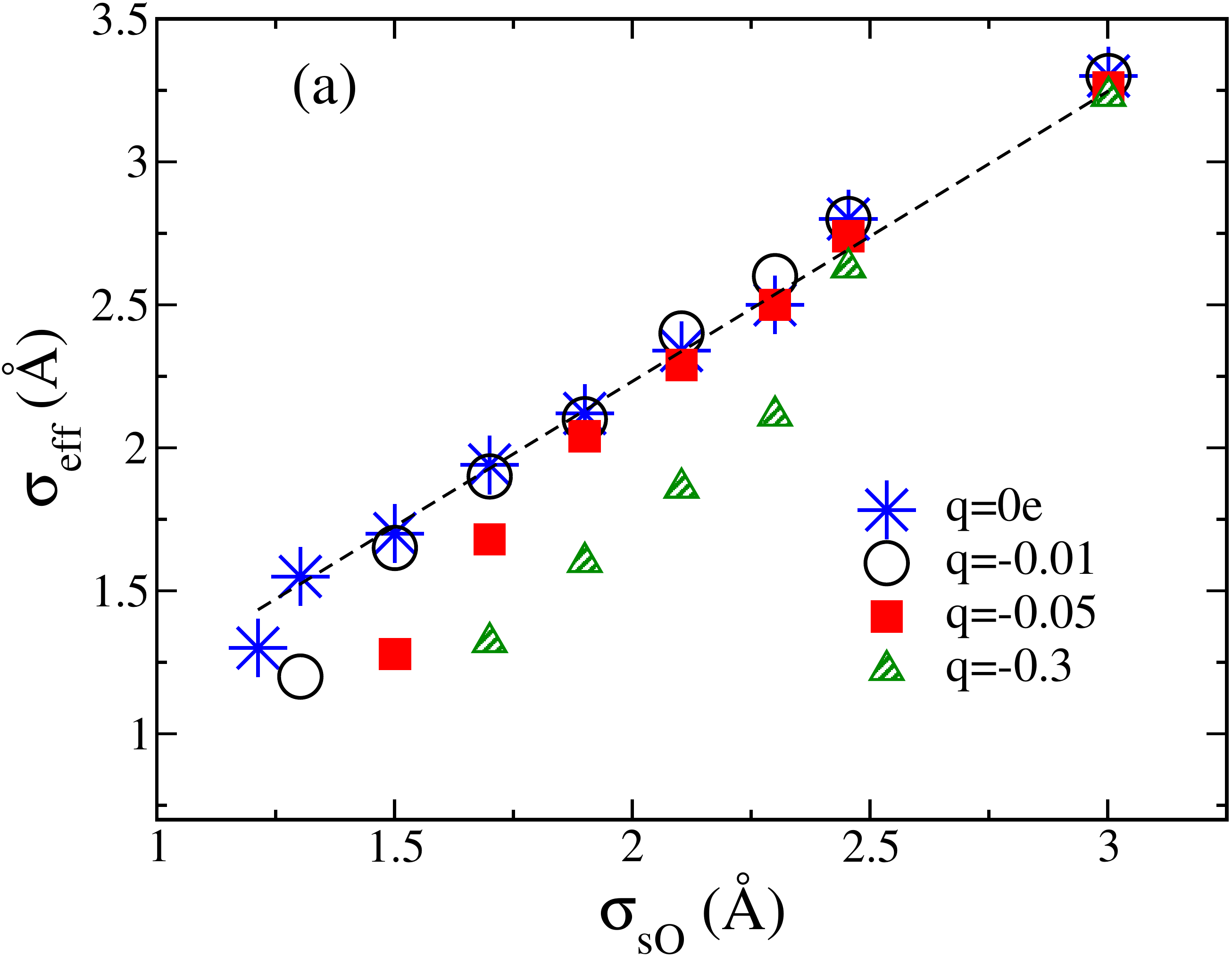}}
	\subfigure{
		\includegraphics[width=0.36\textwidth]{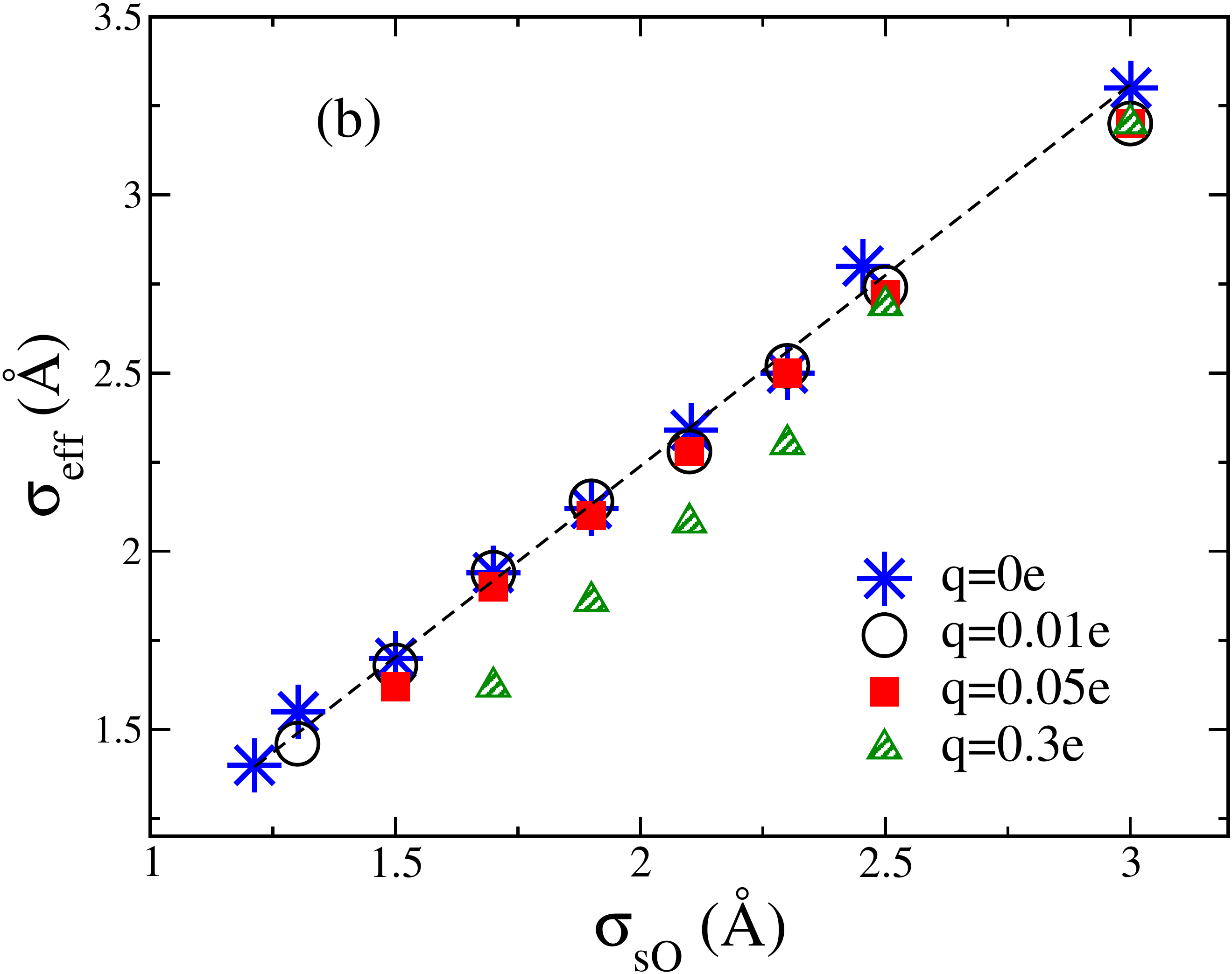}}
	\caption{ \it{The first peak position of the RDF between O and (a)negatively charged and (b)positively charged solutes as a function of solute-solvent diameter $\sigma_{sO}$. Different symbols are used to signify different charges. A dotted straight line is used to highlight the slope followed by the neutral solute. The charged solutes show a deviation from linearity. }}
	\label{fig3}
\end{figure}
\noindent

Ghorai and Yashonath have shown by Voronoi polyhedra analysis that the water structure and thus the water cages around the solute does not change with the change in the charge on the solute\cite{evidence}. Thus in their study they could not explain the shift in the diffusivity maxima in terms of water structure. However, when we plot the water Oxygen and solute (O-SOL) radial distribution function (RDF), we find that the first peak shifts to smaller 'r' value as the charge increases.  
In Fig.\ref{fig3}, for a few systems, specially the ones showing nonmonotonicity, we plot $\sigma_{eff}$ against $\sigma_{sO}$, where $\sigma_{eff}$ is the first peak position of the RDF between O and -ve/+ve solutes.  $\sigma_{eff}$,  defines the first shell or the effective radius of a charged solute atom. We also plot the $\sigma_{eff}$ of the neutral solute as a function of $\sigma_{sO}$ and find that it shows a linear behaviour. We use this linear behaviour as a guide. 
 In this plot, we observe that for charged solutes as $\sigma_{sO}$ decreases, there is a drop/deviation from linearity. The deviation increases with charge and appears at a larger $\sigma_{sO}$ value. Note that, this deviation from linearity of the $\sigma_{eff}$ vs $\sigma_{sO}$ plot appears at the  $\sigma_{sO}$ value where the diffusivity  maximum is present. We also note that, this deviation is more prominent for -ve charges rather than for +ve charges.
The +ve charges feel attracted towards the O and the -ve charges towards H.
The H has a smaller radius than the O and in our model the H is a point particle. This allows more proximity between the -ve charges and the H and thus larger interaction. This large attraction leads to a large reduction in the effective solute-solvent radius. 
Let us consider this change in $\sigma_{eff}$ as a change in the water structure around the charged solutes. Next, we plot D vs 1/$\sigma_{eff}$ (Fig.4) for the negative solutes as the shift is prominent in this case. We find that the diffusivity maxima for all the charged systems still does not appear at the same value of $\sigma_{eff}$.  Thus although the water structure around a solute changes with charge the shift of the diffusivity maxima is not related to that.

 Note that, there is a basic difference in the way the interaction potential changes with size for a charged and a neutral solute.
For neutral LJ solute, the shape of the potential remains same but its range shifts with solute-solvent diameter. In charged system both Coulombic and LJ interactions are present. Note that, the LJ part is size dependent but the electrostatic part is not. This leads to both change in shape and range of the potential.  
It appears that the phenomena we see in Fig.2 is due to the competition between the LJ and the Coulombic interactions. Earlier studies have also made such conclusion\cite{Born1920,chen,boyd,onsager,zwanzig1,zwanzig2}. 
To understand this phenomenon for our set of systems, in Fig.5 we plot the potentials independently arising due to the LJ and Coulombic interactions for negative solutes at three different values of $q_{s}$. 
Our analysis although now focuses on three systems the conclusions are general.

For large sizes, we find the potential is primarily dominated by the LJ part. This is more prominent for small charges, like for q=0.001e and 0.01e. This is precisely the reason the diffusion and $\sigma_{eff}$ values for large sizes are similar to that of neutral solutes (Fig.2a and b). However, as the size of the ionic solute decreases, we see a departure of the diffusion (and also $\sigma_{eff}$) value from that of the neutral 
solute and also the potential starts having contribution from Coulombic interaction. From the plot of the potential we find that the transition from LJ dominated to Coulombic dominated regime takes place at a size where
the diffusivity maxima is present. For higher charges, as the Coulombic interaction is stronger and felt at longer distance, this transition happens at a larger size. 
Note that, in Fig.1(b), for neutral solute, we plot the LJ potential as a function of $\sigma_{sO}$ and do not find any nonmonotonic behaviour which is present in case of charged solutes (Fig.5) although the diffusion has a size dependent maxima. Thus the nonmonotonicity in the potential and its connection with the nonmonotonicity in diffusion as a function of solute size is a feature present only for charged solute system. 

\subsubsection{Reduction of height of diffusivity maxima}
Next, we address the phenomena of the reduction of the diffusivity maxima with increase in solute charge. This effect is just the opposite of that obtained for neutral solutes. 
As mentioned earlier, the diffusion value of larger sizes trace that of neutral solutes. For the neutral solutes at 300K and $\epsilon_{sO}=$1.5846 kJ/mol, the diffusion increases as the size decreases. For larger sizes even for charged solutes we see a similar effect. However, beyond a certain size depending on the charge on the solute there is a shift from LJ dominated to Coulomb dominated regime and a drop in the diffusion value. As, shown earlier the transition shifts to higher sizes with increase in charge. The diffusion value of higher sizes are smaller thus this shift leads to smaller value of diffusion at the maxima.
 Also note that as the charge increases even in the LJ dominated regime i.e, for the large sized solutes the Coulombic interaction is present. The effect of this can also be seen in the diffusion values (Fig.2). The presence of the Coulombic interaction leads to reduction of diffusion value. Thus the difference in diffusion values between the neutral and charged solutes even at large sizes is present and the difference increases with charge.

\subsubsection{SE prediction, cage diffusion and LE}
Next, we analyze the value of diffusion as compared to that predicted by Stokes-Einstein relation. 
In the diffusivity plots we show the SE prediction. The diffusion values of the neutral solutes are much higher than the SE prediction. This has been studied in details by us and other groups \cite{sayantan,manoj}. Since the solute particles not only diffuse via the viscous mode but if size permits also explore the route where they diffuse through the transient solvent cage, it leads to a higher value of diffusion. We find that as we switch on the charge the diffusion of the larger solutes are still higher than the SE prediction.
 This implies that these solute particles also explore the diffusion through the transient water cage. However, even before it reaches the levitation size, due to the dominance of Coulombic attraction, the diffusion value decreases sharply. Thus, although a diffusivity maxima is present in ionic solute diffusion and for certain sizes and for small charges, the solute explores transient cage diffusion, the origin of the maxima appears to be different from the argument given for LE.

\begin{figure}[ht]
	\centering
	\includegraphics[width=0.45\textwidth]{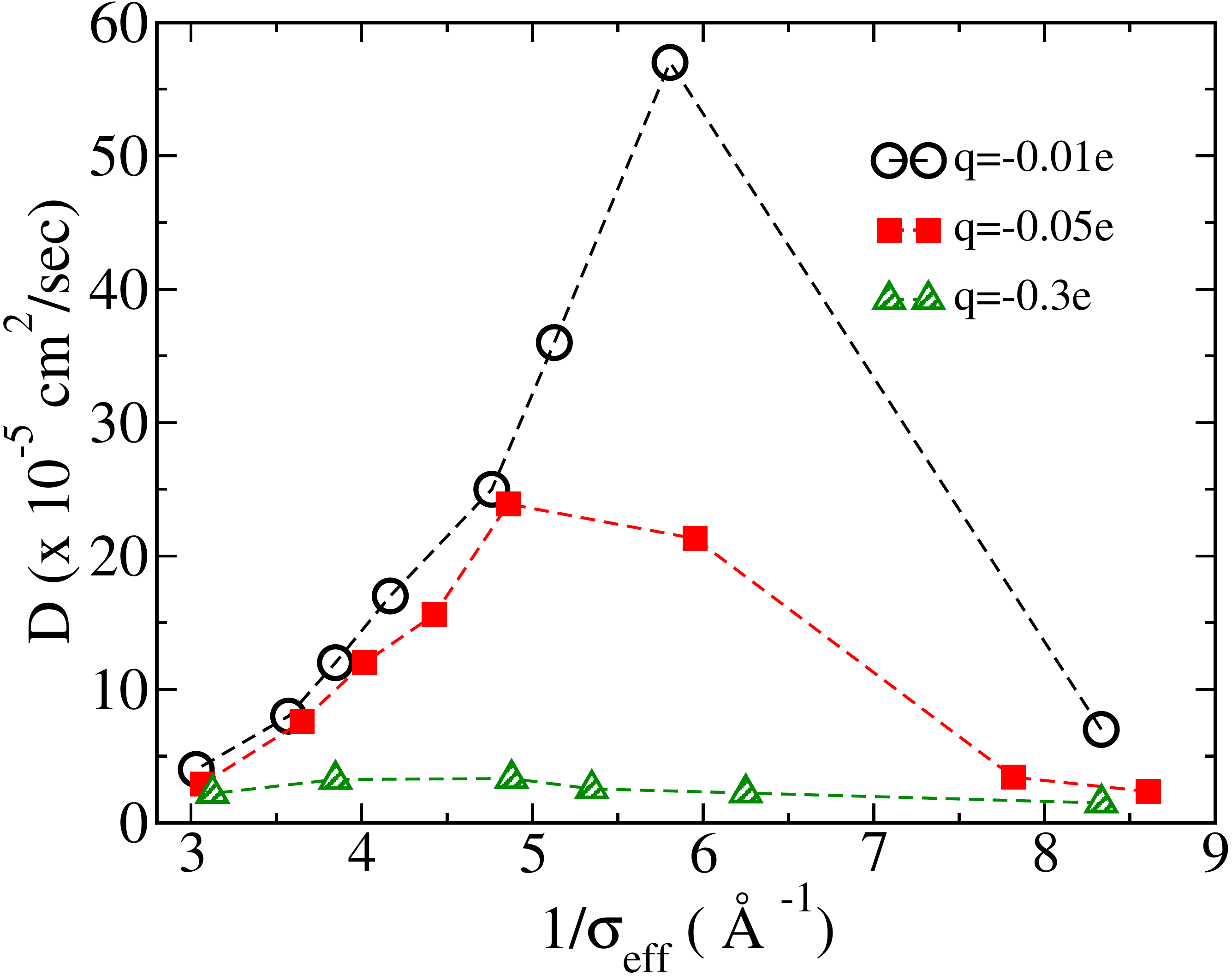}
	\caption{ \it{The diffusion coefficient D of the solute particles of different charges as a function of the first peak position of the RDF between O and negatively charged solutes showing diffusivity peaks at similar position.}}
	\label{fig4}
\end{figure}
\noindent

\begin{figure*}[ht]
\centering
\subfigure{
	\includegraphics[width=0.36\textwidth]{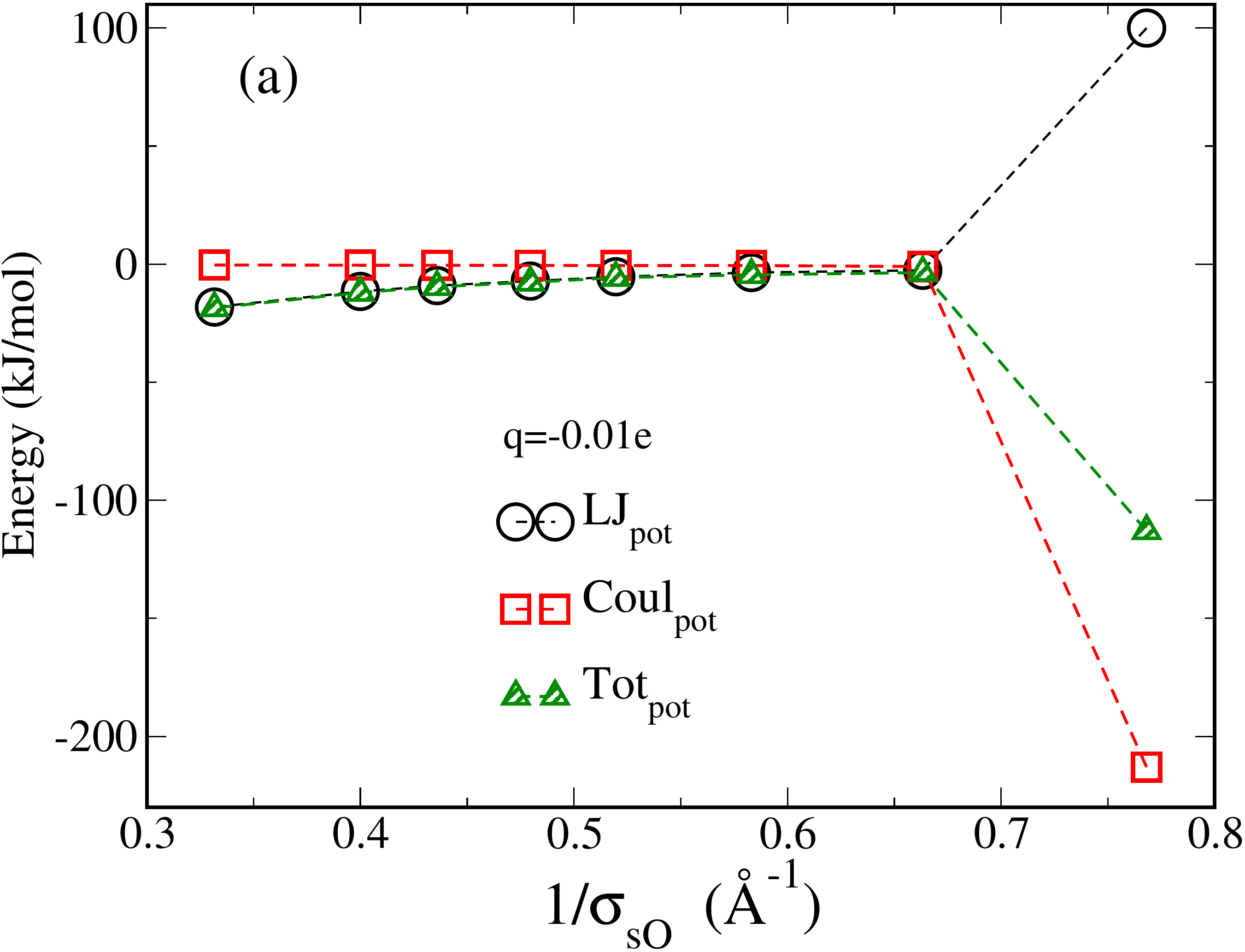}}
\subfigure{
	\includegraphics[width=0.36\textwidth]{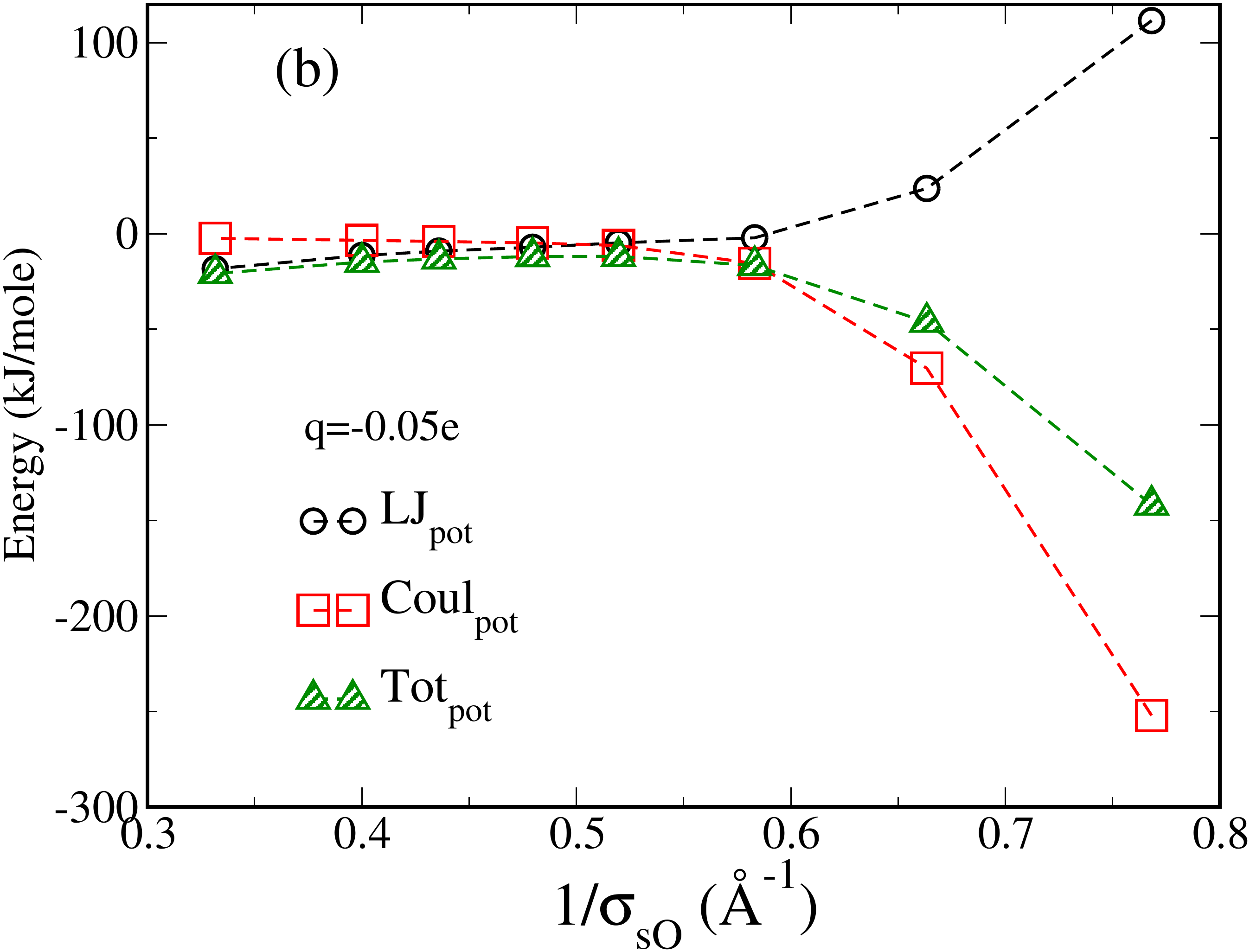}}
\subfigure{
	\includegraphics[width=0.36\textwidth]{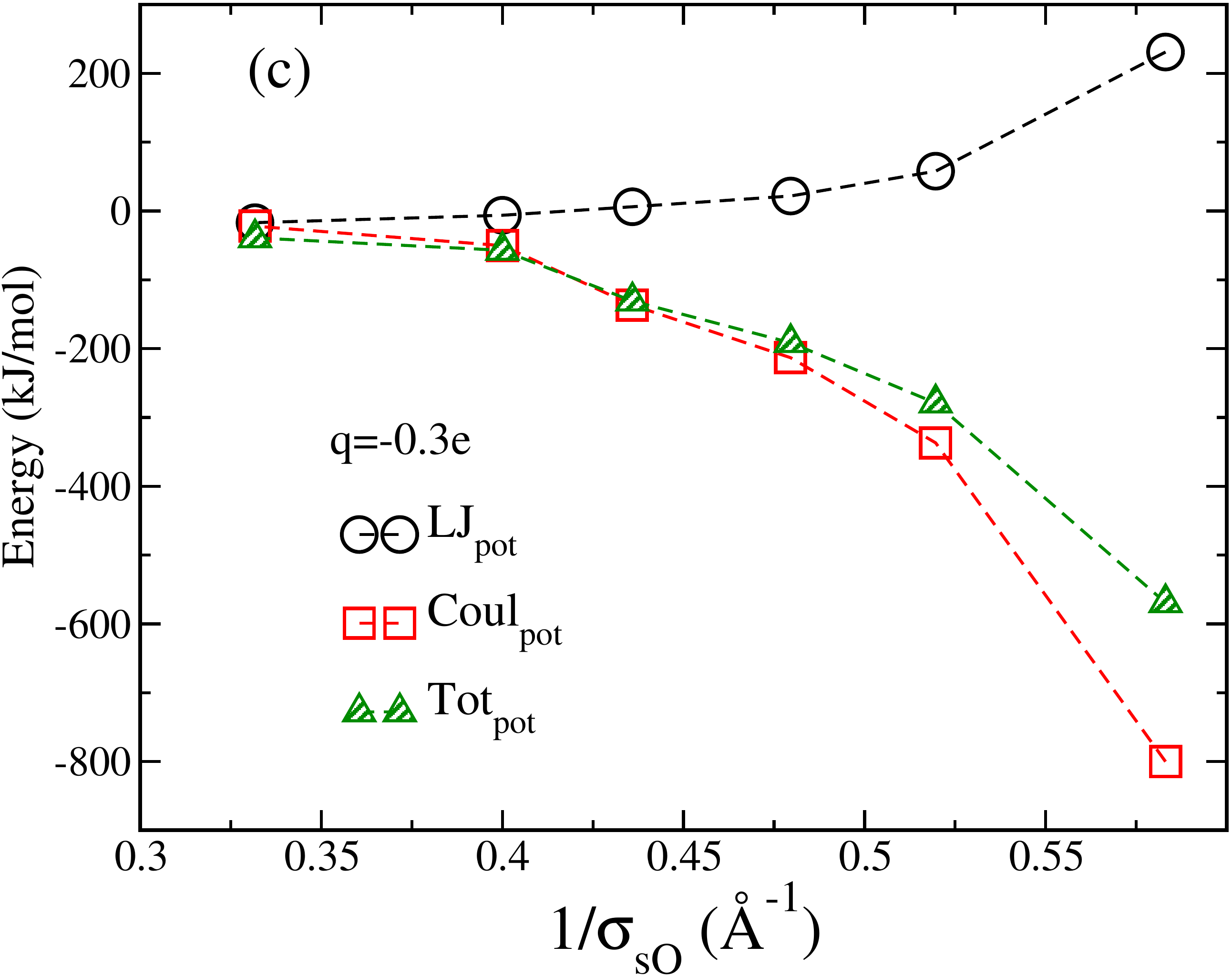}}

\caption{\it{ Individual contributions from the LJ, Coulomb and total potential for three different charges are plotted against 1/$\sigma_{sO}$. (a)$q_{s}$=-0.01e, (b)$q_{s}$=-0.05e and  (c)$q_{s}$=-0.3e.
		For $q_{s}$=-0.01e and at larger sizes the LJ potential dominates and at smaller sizes the Coulombic potential dominates. As the charge increases ($q_{s}$=-0.3e) even at higher sizes we find contribution from Coulombic interaction. The transition from LJ dominated to Coulomb dominated regime gives rise to the diffusivity maxima.}}
\label{fig6}
\end{figure*}
\noindent

\section{Conclusion}

In this work, we do a comparative study of the size dependence of diffusion 
for both charged and neutral solutes in water. Earlier studies by
Yashonath and coworkers \cite{padma,evidence} have reported that 
both the systems show a nonmonotonicity in the size-dependence
of diffusion and have connected it to the LE usually obtained for neutral solutes \cite{manoj,size,Bhatia,rajappa}.
According to LE for attractive solute-solvent interaction, when 
the solute diffuses through the transient solvent cage, only for certain size of the solute
in comparison to the neck of the cage
there is a force balance and this leads to a diffusivity maximum. 
The smaller solutes feel an uneven attraction towards one direction, which
slows down the solute motion and leads to reduced diffusion. 
Ghorai and Yashonath have claimed that similar phenomenon is present even for charged solutes \cite{evidence}.
However, traditionally the diffusivity maximum in ionic systems is explained as an effect of the interplay between hard sphere repulsion and ionic attraction\cite{Born1920,chen,boyd,onsager,zwanzig1,zwanzig2,ranjit,bagchi,biswas,rasaiah,wolynes}.

In case of neutral solutes, we find that, there exist a diffusivity maxima
in the size dependence of diffusion.
The diffusion value at the maxima is independent of the strength of attraction, and the relative height of the maxima 
increases with strength of attraction. 
For charged solutes, we find that for certain values of  
the charge, there exists a diffusivity maxima. 
The position of the maxima shifts to larger sizes as the interaction strength (charge) increases. Also,
the relative height of the maxima decreases with interaction strength. Both these phenomena are different from what 
we find in case of neutral solutes.
By analyzing the RDF, we show, that the structure of the solvent around the charged solute changes with the charge.
However, this change in structure cannot explain the shift of the diffusivity maxima.
We find that large sized solutes at smaller value of charge behave like neutral solutes. Thus the diffusion value of the former is quite close to that of the latter. For neutral solutes, as the value of diffusion grows with the decrease in size so does that for charged solutes.
However, as the size decreases there is a transition in the interaction potential from Lennard-Jones dominated to Coulomb dominated regime where the latter is strongly attractive and with this, there is a drop in the diffusion value. This gives rise to the nonmonotonicity in size dependence of diffusion. This transition to Coulomb dominated regime shifts to higher sizes as the charge on the solute increases. Since the value of the diffusion is smaller at higher sizes thus the shift of the transition to higher sizes leads to the reduction of the height of the diffusivity maxima with the increase in charge.  

We also show that for lower charges, closer to that of neutral solutes the value of the diffusion at higher sizes are much larger than that predicted by the SE relation. Earlier studies on neutral solutes have shown that this high diffusion value arises as the solute not only diffuses via the viscous mode but also exploits cage diffusion \cite{sayantan}. This implies that these charged solutes also explore the cage diffusion. However, even before we reach the size where the solute can exploit the cage diffusion the most by levitating through the cage, the Coulombic interaction starts dominating and there is a decrease in the diffusion value.
Thus we show that although both charged and neutral solutes show a diffusivity maxima, the origin of it is quite different. In this work we also clearly show that as discussed in earlier literatures \cite{Born1920,chen,boyd,onsager,zwanzig1,zwanzig2,ranjit,bagchi,biswas,rasaiah,wolynes}, for charged solutes the diffusivity maxima is an effect of the interplay between hard repulsion and Coulombic attraction.

\section{References}

\end{document}